\documentclass[epjH]{svjour}
\usepackage{graphicx}
\begin{document}
\title{Century of $\Lambda$}
\author{Bohdan Novosyadlyj\inst{1,2}\fnmsep\thanks{\email{bnovos@gmail.com}}}
\institute{Astronomical Observatory of Ivan Franko National University of Lviv, Kyryla i Methodia str., 8, Lviv, 79005, Ukraine \and International Center of Future Science of Jilin University, Qianjin Street 2699, Changchun, 130012, P.R.China}
\abstract{The cosmological constant was proposed 100 years ago in order to make the model of static Universe, imagined then by most scientists, possible. Today it is the main candidate for the physical essence causing the observed accelerated expansion of our Universe. But, as well as a hundred years ago, its nature is unknown. This paper is devoted to the story of invention of $\Lambda$ by Albert Einstein in 1917, rejection of it by him in 1931 and returning of it into the great science by other scientists during the century. The aim is to once again emphasize prominent role of cosmological constant in the development of ideas of modern physics and cosmology, focusing on the main points and publications, the choice of which may have a certain part of subjectivity. 
} %end of abstract
\maketitle
 \section{Introduction}

The fundamental physical constants always attract the attention of physicists due to their importance for agreement of theoretical predictions with corresponding experimental data or astronomical observations. Most of them have definitive physical interpretations and conventional values up to the experimental uncertainties of their determination using the advanced methods and instrumentation. But one of them -- the cosmological constant -- is the subject of hot discussions about its necessity, value and physical essence during the last century. At the beginning of 1917 Albert Einstein's paper in Sitzungsberichte der Koeniglich Preussischen Akademie der Wissenschaften (Berlin) intrigued the physical community and Wilem de Sitter's paper in Monthly  Notices of the  Royal Astronomical Society at the end of that year attracted the attention of astronomers. In this paper, devoted to the century of cosmological constant, we briefly review the motivation of its including in general relativity, using it for interpretation of the cosmological observations and the unsuccessful attempts to give it some fundamental meaning. The comprehensive historical review of the story of the cosmological constant one can find in the recent papers \cite{Cormac2017,Cormac2018} as well as the complete list of publications devoted to it.

\section{``Cosmological Considerations on the General Theory of Relativity''} 

The title of this section is the English version of title of Albert Einstein's paper published in Proceedings of the Royal Prussian Academy of Sciences (Berlin) on 15th February 1917 \cite{Einstein1917}, in which the cosmological constant $\lambda$ in the equations of general relativity\footnote{Here and below the notation of original papers is kept.} has been introduced:
\begin{eqnarray}
&G_{\mu\nu}-\lambda g_{\mu\nu}=-\kappa\left(T_{\mu\nu}-\frac{1}{2}g_{\mu\nu}T\right),& \nonumber\\  \label{ee1}\\
&G_{\mu\nu}=-\frac{\partial}{\partial x_{\alpha}}\left\{\begin{array}{cc} \mu & \nu \\ \multicolumn{2}{c} \alpha\end{array}\right\} + \left\{\begin{array}{cc} \mu & \alpha \\ \multicolumn{2}{c} \beta\end{array}\right\}\left\{\begin{array}{cc} \nu & \beta \\ \multicolumn{2}{c} \alpha\end{array}\right\}+\frac{\partial^2\log{\sqrt{-g}}}{\partial x_{\mu}\partial x_{\nu}}-\left\{\begin{array}{cc} \mu & \nu \\ \multicolumn{2}{c} \alpha\end{array}\right\}\frac{\partial\log{\sqrt{-g}}}{\partial x_{\alpha}}&. \nonumber 
\end{eqnarray}
Here and below $g_{\mu\nu}$ is the metric tensor of Riemannian spacetime in coordinates $x_1$, $x_2$, $x_3$, $x_4$ defining the square of interval between infinitesimally close events as $ds^2=g_{\mu\nu}dx^{\mu}dx^{\nu}$,  $\left\{ \begin{array}{cc} \mu & \nu \\ \multicolumn{2}{c} \alpha  \end{array}\right\}$  are  coefficients of the  Riemannian connection or Christoffel's symbols of the second kind (other notation $\Gamma^{\alpha}_{\mu\nu}$), $T_{\mu\nu}$ is the energy-momentum tensor of matter and fields, $\kappa$ is Einstein gravitational constant. 
Adding/subtracting the term $\lambda g_{\mu\nu}$ does not change the general covariance of equations, does not break the energy and momentum conservation law. For the sufficiently small value of constant $\lambda$ the predictions of theory based on the equations (\ref{ee1}) are compatible with observations in the Solar System. For Einstein the motivation for introducing it was the incompatibility of the equations of general relativity from 1916 year with the model of homogeneous isotropic static spatially closed (positive curvature of the 3-space) world. The justification of such a model makes the most of text of the paper -- six pages out of ten (sections 2-3), while the main mathematical part -- equations (1), their analysis and solution -- constitutes incomplete two journal pages (sections 4-5). 

From the absence of observable data on significant redshift or blueshift of stars he makes conclusion that the world is static (page 152, last sentence\footnote{In it, Einstein uses ``quasistatische'' - quasi-static, while in the entire text of the paper he uses ``static''.}) and he assumes the same energy-momentum tensor for medium of stars as for dust-like matter 
\begin{equation}
T_{\mu\nu}=diag\{0,\,0,\,0,\rho\}, \label{tei}
\end{equation}
where $\rho$ is the matter density. For the world with closed 3-space Einstein assumes the metric of 3-sphere with radius $R$, which is in the imaginary Euclidean 4-space: 
\begin{equation}
g_{\mu\nu}=-\left(\delta_{\mu\nu}+\frac{x_{\mu}x_{\nu}}{R^2-(x_1^2+x_2^2+x_3^2)}\right). \label{g}
\end{equation}
Equation (1) is satisfied for (\ref{tei}) and (\ref{g}) if the following equalities holds:
\begin{equation}
\lambda=\frac{\kappa\rho}{2}=\frac{1}{R^2}. \label{me}
\end{equation}
They mean that $\lambda$ is positive as well as the curvature of 3-space, since the density of matter is positive. The last formula -- numbered (15) -- in the paper is the expression for full mass of the Universe, which is determined by one of these three values, 
\begin{equation}
M= 2\pi^2 R^3\rho =4\pi^2\frac{R}{\kappa}=\frac{\sqrt{32}\pi^2}{\sqrt{\kappa^3\rho}}. \label{mass}
\end{equation}
This equality can be continued as $...=\frac{4\pi^2}{\kappa\sqrt{\lambda}}$. 
Such model of the world is now called the Einstein one -- homogeneous static closed universe of finite mass, the value of which determines positive value of the curvature and cosmological constant. Or -- on contrary -- the cosmological constant determines the mass and 3-curvature of the world.  

Also for Albert Einstein the interpretation of cosmological constant in the Newtonian approximation of gravity was important; that is why the paper begins with discussion of the problem of Newtonian cosmology -- infinity of the gravitational potential on the spatial infinite boundary of homogeneous static world (introduction and section 1). He shows that adding to the left-hand side of Poisson equation the term\footnote{Here Einstein notes that this analysis should not be taken too seriously -- only as an introduction to the main part.} ``$-\lambda\phi$'' ensures finiteness of the gravitational potential $\phi$ and inviolability of the law of universal gravitation in the vicinity of massive bodies, where the gradient of background value of the gravitational potential is negligibly small. Such a world is infinite, its average density $\rho$ is constant and small, the gravitational potential on spatial scales significantly larger than the size of stars is constant and is determined by the mean density and the value of the cosmological constant: $\phi=-4\pi K\rho/\lambda$. Here $K$ is Newtonian gravitational constant.
Harvey and Schucking (2000) (see also \cite{Cormac2017,Cormac2018} and citing therein) have shown, however, that the additional term  ``$-\lambda\phi$'' in Poisson equation does not agree with Newtonian approximation of eq. (\ref{ee1}), the term ``$c^2\lambda$'' follows from it. Whether was it Einstein's mathematical mistake?  In my opinion, it was more important for him to find excuses for manual correction of equations for the beauty of the model of static closed world, which he received in this work. He does not discuss the Newtonian approximation of such model because the world as whole is subject of discussion. 

During the next decade at first theoretically by Alexander Friedmann (1922) and Georges Lema\^{\i}tre (1927) and later observationally by Edwin Hubble (1929) who determined the distances to galaxies with measured redshifts the nonstationarity of homogeneous isotropic world has been proven, nonetheless, the introduction of cosmological constant by Einstein in this way can be considered an outstanding scientific guess. In fact, which observational arguments for the homogeneity and isotropy of the world, its static or closed nature were available that time? The visible distribution of stars in the space is heterogeneous and anisotropic because the Solar System is inside the stellar system Milky Way and is far from its center. Einstein knew this well. Other stellar systems -- galaxies -- have entered into scientific picture of the world much later. In this paper Einstein argues homogeneity, staticity and closeness of the world using the fact of smallness of the relative speeds of stars in comparison with the speed of light. Now we know that this is not a reliable argument.  

\section{De Sitter, Friedmann, Lema\^{\i}tre cosmological models and Hubble law}

Wilem de Sitter, who ``made a significant contribution to solving an important cosmological problem about the space-time structure''\footnote{A. Einstein, New York Times, Nov. 22, 1934}, was the first who supported the idea of cosmological constant: on 31 March 1917 in Proceedings of the Royal Academy of Sciences of the Netherlands he has published the paper ``The relativity of inertia. Notes on Einstein's latest hypothesis'' \cite{deSitter1917a}, in which he analysed the geometric properties of Einstein's model of the world and model of the world with cosmological constant $\lambda$ but without matter ($\rho=0$). In November of the same year in Monthly Notices of the Royal Astronomical Society has he published the paper ``About the theory of gravitation of Einstein and its significance for astronomy. The third article'' \cite {deSitter1917b}. In this paper de Sitter developed the model of the  world with cosmological constant. At the beginning he showed that the equation (\ref{ee1}) can be obtained from the generalised Hamiltonian principle if we assume
\begin{equation}
H_3=\int\sqrt{-g}(G-4\lambda)d\tau. \nonumber
\end{equation}
Thus, introduction of $\lambda$ leaves all the conservation laws that follow from this principle valid. Further, he shows that since the curvature of 4-space is proportional to $G$, even in the case of empty universe ($T=0$) the space is curved: $R^{-2}=\lambda/3$. De Sitter initially did not realise that its coordinate system was not comoving. Only later, after the Lema\^{\i}tre's paper of 1927 and the discovery of expansion of the Universe by Edwin Hubble in 1929 he remarked that in the comoving frame with cosmological time the model is non-stationary: $R\propto\exp{\sqrt{\lambda/3}t}$ \cite{deSitter1930}. 
Such model of the world is called de Sitter model. In the 1970s it played a key role in development of the inflationary scenarios of the early Universe, and is now seen as prediction for the future expansion if the dark energy is cosmological constant or something similar to it.  

Important contribution to the analysis of Einstein's cosmological equations (\ref{ee1}) was made by Alexander Friedmann in his works ``The curvature of space'' (1922) \cite{Friedmann1922} and ``The possibility of a world with a constant negative curvature of space'' (1924) \cite{Friedmann1924}. He analysed the equations (\ref{ee1}) in the metric 
\begin{equation}
ds^2=-\frac{R^2(x_4)}{c^2}\left(dx_1^2+\sin^2x_1dx_2^2+\sin^2x_1\sin^2x_2dx_3^2\right)+dx_4^2, \nonumber
\end{equation}
in which they are reduced to two ordinary first and second order differential equations\footnote{They are now called the Friedmann equations.} 
\begin{equation}
\frac{R^{\prime 2}}{R^2}+\frac{2RR^{\prime\prime}}{R^2}+\frac{c^2}{R^2}-\lambda=0, \quad \quad
\frac{3R^{\prime 2}}{R^2}+\frac{3c^2}{R^2}-\lambda=\kappa\rho, \label{fre}
\end{equation}
where prime denotes the derivative with respect to the time coordinate $x_4$. He has shown that in addition to the static Einstein and de Sitter solutions (conditions $R^{\prime}=R^{\prime\prime}=0$) the non-stationary solutions of equations that describe the monotonous expansion, monotonous compression or oscillations also exist. They are determined by the initial conditions, the values of cosmological constant,  density of matter and  curvature of 3-space at certain time that ``passed from the beginning of the world''. The title and topic of the second paper are caused by the fact that in Einstein model (\ref{me}) the curvature of 3-space as well as the cosmological constant can have only positive values, since so does the matter density. De Sitter also analysed only the spaces with positive curvature, perhaps because this is required by Einstein model. 
Friedmann has shown that the world with negative 3-curvature can be static for the negative value of cosmological constant when the density of matter is negligible, while the non-stationary world may have any non-zero positive value of matter density. These two Friedmann's papers initiated completely new look on the world -- non-stationary world with arbitrary curvature of 3-space, positive density of matter and indefinite cosmological constant. Discussing a possible variant of the model of our Universe, he notes that ``in the formulas we have obtained the cosmological value $\lambda$ is not defined, being an unnecessary constant of the problem ''\cite{Friedmann1922}. Perhaps therefore Einstein's first reaction to Friedmann's paper was rejection of it, and only later he admitted that ``Friedmann's results are correct and shed the new light. It turns out that the field equations allow both static and dynamic solutions of the quantities that determine the structure of the space'' \cite{Einstein1923}.

Independently of Friedmann the non-stationary solutions of cosmological equations (\ref{ee1}) were obtained by Georges Lema\^{\i}tre in 1927 in the paper ``A homogeneous universe of constant mass and increasing radius accounting for the radial velocities of extra-galactic nebulae'' \cite{Lemaitre1927}. In addition, he obtained the expression for redshift of the sources of light in the Universe that expands and discovered the law of moving apart of galaxies, which is now called the Hubble law:  
\begin{equation}
\frac{v}{c}=\frac{R_2}{R_1}-1=\frac{R^{\prime}}{R}r \label{Llow}
\end{equation}
He assumed further that the extragalactic nebulae have the same brightness and estimated the distances to 43 nebulae with the determined radial velocities 
\cite{Stromberg1925} using the known formula that links the distance to the apparent magnitude: $\log{r}=0.2m+4.04$. This allowed him to estimate magnitude of the rate of expansion of the Universe using (\ref{Llow}): 
\begin{equation}
\frac{R^{\prime}}{R}\approx630 \frac{\rm{km}}{\rm{c\cdot Mpc}}.\label{LH_0}
\end{equation}
Since the paper was published in French, it was unnoticed by Edwin Hubble, who in 1929 \cite{Hubble1929} by comparing the velocities of galaxies determined from the Doppler effect with the distances to them found the correlation between them -- the proportionality of velocity to distance -- $v=Kr$, where the coefficient of proportionality\footnote{Now it is denoted by the letter $H$ and is called the Hubble constant.} $K$ describes the rate of expansion of the Universe $K=R^{\prime}/R$. He determined its magnitude based on data about the velocities and distances to 24 galaxies 
\begin{equation}
K\approx465\pm50 \frac{\rm{km}}{\rm{c\cdot Mpc}}.\label{HH_0}
\end{equation}
Here the distances to galaxies were determined assuming that the brightest stars in galaxies have the same luminosities. The motion of Sun in the Galaxy was also taken into account. This became the confirmation of non-stationary model of the Universe and brought to Hubble the glory of discoverer of its expansion. The Lema\^{\i}tre's method, based on the incorrect assumption about the same luminosities of all galaxies, gave nevertheless similar value of the rate of expansion of the Universe. However, in the reprint of the paper in English in Monthly Notices of the Royal Astronomical Society in 1931 \cite{Lemaitre1931} he removed himself the description of rate of expansion, which is evidenced by correspondence of the author with the journal editors, according to the recent studies of this story  \cite{Livio2011}. 

\section{Einstein's denial of $\lambda$}

In the paper ``On the cosmological problem of the general relativity'' Einstein notes that the solution (\ref{me}) is obtained from the Friedmann equations (\ref{fre}) for $R$ which is constant in time and space \cite{Einstein1931}. ``However with the help of these equations, one can show that this solution is unstable. It means, that any solution, which at some moment of time is slightly different from the static, will over time be more and more different from it. Already for this reason, without telling as for the results of Hubble's observations, I do not consider it possible to attribute the physical meaning to my previous solution. In this connection one can ask whether it is possible to describe the observation without introducing a $\lambda$-term that is clearly contradictory from the theoretical point of view'' \cite{Einstein1931}. In this paper and in the next one with de Sitter published in 1932 \cite{Einstein1932} he showed that it is possible. In addition to the second edition of ``The Essence of the Theory of Relativity'' published in 1945 Einstein wrote: ``The introduction into the gravitational equations of the cosmological term is possible, albeit in terms of the theory of relativity is not logically necessary. As Friedmann showed for the first time, the density of matter that is finite everywhere can be reconciled with the first form of equations of gravity, assuming that the metric distance between two material points changes over time. Already one requirement of spatial isotropy of the Universe leads to the scheme of Friedmann. There is no doubt that this is the most general scheme that gives a solution to the cosmological problem'' \cite{Einstein1945}. Einstein could not get rid of the ``feeling of guilt'' that he has introduced into physics the cosmological constant as an ``extra essence'', so he occasionally returned to this topic. In the letter to Lema\^{\i}tre, written on 26 August 1947, we find: ``Since I introduced this constant, I was accompanied by a sense of unclean conscience. ... I believe that it is actually very ugly ... and I can not believe that such ugly thing could have been realised in the nature'' \cite{Kraph2017}. George Gamow wrote in the book ``My World Line''  that during one of his conversations with Einstein in Princeton he said ``that the introduction of the cosmological term was the biggest blunder he ever made in his life''. Gino Segre described also this confession in the book ``Ordinary Geniuses'', naming one of 45 sections as ``Einstein's Biggest Blunder''. However, the history of $\lambda$-term did not end there. It was destined to have a long and interesting ``life''. Steven Weinberg noted in his book ``Cosmology'' published in 2008: ``Einstein's mistake was not that he introduced the cosmological constant -- it was that he thought it was a mistake'' \cite{Weinberg2008}.  The following sections of this paper confirm this.

\section{$\Lambda$: vacuum, inflation, clusters of galaxies and quasars }

Despite Einstein's scepticism about the possibility of realisation of cosmological constant in the nature, it continued to live its own life in the minds of physicists and cosmologists. Georges Lema\^{\i}tre, Richard Tolman, Willem de Sitter, Arthur Eddington and other remained its supporters (for more details see \cite{Cormac2018}). In 1934 G. Lema\^{\i}tre has suggested that since the energy associated with it is Lorentz invariant, then the cosmological constant is the vacuum energy density \cite{Lemaitre1934}. But he has not linked it with the zero-point energy of vacuum, although this idea\footnote{The history of the idea and its development are described in the paper \cite{Kragh2012}.} was actively discussed at that time. The Heisenberg uncertainty principle and quantum field theory are today the theoretical basis for understanding of the physical nature of vacuum. Despite the significant quantitative differences between the allowed value of energy density related to the cosmological constant and the estimates of value of the zero-point energy of vacuum (see next paragraph), this idea has become attractive for many physicists. In 1965 in the paper ``The algebraic properties of the energy-momentum tensor and the vacuum-like state of matter'' Erast  Gliner\footnote{He was born in Kyiv in 1923, as a student survived the blockade of Leningrad, fought in the World War II, lost his hand, survived to pass through the hell of Stalin prison camps and later to propose the key idea of inflationary models of the Universe. Now, he is almost forgotten at motherland, lives in California.} proposed the physical interpretation of cosmological constant that solved the problem of cosmological singularity and initial state of the Universe \cite{Gliner1965,Gliner1970}. In accordance with this hypothesis, initially there was vacuum described by the cosmological constant and de Sitter model. It gave birth to the matter which began to expand due to the ''cosmic repulsion`` of such vacuum. In such a way the matter and the observed cosmological expansion of the Universe arose. In fact, this was the first scenario of cosmological inflation developed later by A. Starobinsky (1979) \cite{Starobinsky1979}, A. Guth (1981) \cite {Guth1981}, A. Linde (1982) \cite{Linde1982}, A. Albrecht and P. Steinhard (1982) \cite{Albrecht1982} to modern scenarios with the quantum one-loop corrections or the inflaton field as physical basis of inflation. In different years they all deservedly received prestigious awards for the development of inflationary models of the Universe ... except for E. Gliner.  

Another interesting application of the cosmological constant was suggested by Georges Lema\^{\i}tre in 1933 \cite{Lemaitre1933} (see also \cite{Lemaitre1934}), namely, he suggested that the cluster of galaxies is static formation, the size of which is determined by the equality of forces of the cosmic repulsion, which is described by the cosmological constant, and the gravity of matter. In the same year Fritz Zwicky published his paper \cite{Zwicky1933} about the cluster of galaxies in the Coma constellation, in which he proved using the virial theorem that a significant part of its mass is dark. This was the first indication of existence of the dark matter.

At the end of 1960s the interest in cosmological constant increased because of detection of the plateau in redshift distribution of quasars and the unusual relationship between luminosity, apparent stellar magnitude and distance to them calculated in the models without cosmological constant. In the papers \cite{Petrosian1967,Shklovsky1967,Kardashev1967} it was shown that they are easily explained in the Lema\^{\i}tre models of 1927. In the paper by J. Shklovsky the cosmological constant is marked by the capital Greek letter ''Lambda``, what is now commonly accepted, and the paper by N. Kardashev is called ''Lema\^{\i}tre's Universe and Observations``. In such a way the soviet scientists celebrated the 50th anniversary of cosmological constant and paid tribute to Georges Lema\^{\i}tre who passed away a year before the publishing of these papers. On basis of the data on redshift distribution of quasars and the assumption about their uniform luminosity in \cite{Kardashev1967} the value $\Lambda=4.3\cdot10^{-56}$ cm$^{-2}$ was obtained, which is quite close to the modern determination of it. These works attracted the attention of Y. Zeldovich -- in 1968 in the journal ''Advances in Physical Sciences`` he published the paper ''Cosmological Constant and Particle Physics``  \cite{Zeldovich1968}. In it he related cosmological constant with the zero-point energy of vacuum and the particle physics. In particular, he argues that ''field theory with relativistic-invariant regularisation does not require zero-energy vacuum, but, on the contrary, naturally leads to a situation described by the cosmological constant`` \cite{Zeldovich1968}. However, his estimate of the value of $\Lambda$ on basis of the  particle physics and the dimensional constant method proved to be many orders of magnitude larger than the upper limit of its value which is still allowed by the astronomical observations. This was the first indication of the problem of physical interpretation of $\Lambda$.

\section{Problems of physical interpretation of $\Lambda$}

Modern observable data indicate that $\Lambda\le1.6\cdot10^{-56}$ cm$^{-2}$. The numerical value is derived from measurements of the Cosmic Microwave Background (CMB) anisotropy, the spatial distribution of galaxies, the apparent magnitude-redshift relation for supernovae type Ia and other data giving the value of dark energy density with a few percent accuracy. The cosmological constant can provide the whole density of dark energy or some part of it, therefore $\rho_{\Lambda}\le\rho_{de}$ and $\Lambda=\kappa\rho_{\Lambda} = 8\pi G\rho_{\Lambda}/c^2$ (see next section). Such a small magnitude of cosmological constant is the problem for giving it the physical interpretation. 

Indeed, the original interpretation of cosmological constant provided by Einstein was as the world constant. The equations of general relativity contain also the gravitational constant $G$ and the speed of light $c$. Since in usual units all of them have different dimensions, it is possible to compare their values only in natural (Planck) units, in which $G=c=1$, while $\Lambda\le10^{-122}$! Can any physical theory explain such relationships between its constants? No. In this interpretation the empty space is curved, that violates the basic postulate of general relativity -- the energy-momentum tensor of matter defines the curvature of spacetime. To avoid such violation physicists began to write the cosmological constant on the right-hand side of equations, referring to its connection with the energy, and not with the geometric part of equations. In such presentation the empty space has zero energy-momentum tensor on the right-hand side, and hence zero curvature on the left one.  

Attractive hypothesis about cosmological constant as density of the zero-point energy of vacuum has serious quantitative disagreement with predictions of the standard model of particle physics. Indeed, if we cut the upper limit in integral for the zero-point energy of vacuum (VIII.1) from \cite{Zeldovich1968} at the Planck energy scale, then we find that the density of the zero-point energy of vacuum is $\sim10^{93}$ g/cm$^3$. It corresponds to $\Lambda\sim10^{66}$ cm$^{-2}$, which is 122 (!) orders larger than the allowed value. In the world with such value of cosmological constant there is not enough time to form any structure of the cosmic scale: if it is positive, then such universe will expand too fast, and if it is negative, then it will collapse too early. Zeldovich's idea about the possibility of reducing the zero-point energy of vacuum by ordering of the polarised vacuum \cite{Zeldovich1968} has not been elaborated to the expected level of consistency. So far, it has been possible to reduce the discrepancy from 123 orders of magnitude to 56 \cite{Koksma2011}. The problem remains, but this is not so much a problem of cosmological constant as a problem of quantum field theory. In this context, it is worth mentioning the long-time scepticism of Wolfgang Pauli about reality of the zero-point energy of vacuum in free space. As an argument he cited the estimate of radius of the world in Einstein model: with the energy of vacuum obtained from that-time data on elementary particles the radius of world would never reach the radius of Moon orbit \cite{Kragh2012}.     

Another problem of cosmological constant is as follows. If the cosmological constant corresponds to the density $\rho_{\Lambda} = c^2\Lambda/8\pi G\approx6\cdot10^{-30}$ g/cm$^3$, then it was the same in the Planck epoch, when the energy density of thermal radiation and all other fields reached the Planck density of $\sim10^{93}$ g/cm$^3$. How could the components of the Universe with the 122 orders of magnitude ratio of energy densities co-exist? If, on contrary, it is assumed that all components of the observed Universe have been created by the decaying of initial field, the inflaton, then how the $\Lambda$ component with the energy density 122 orders of magnitude smaller than the energy density of radiation was produced? If it was smaller not by 122 but 120 orders of magnitude, then the Universe would be completely different and we likely would not be in it. The discrepancies become somewhat smaller  ($\sim100$ orders) if we assume that the inflaton has decayed later, at the time scale that corresponds to the energy of grand unification $t_{GUT}\sim10^{-32}$ s, not Planck period  $t_{Pl}\sim10^{-43}$ s. However, the discrepancy of 100 orders is equally impressive and is beyond the scope of description in the language of current physics. This problem is called the fine tuning problem.   

Another problem of cosmological constant is the problem of coincidence: in the modern era its density is close to the density of matter, during the reionization it was close to the density of thermal radiation \cite{Lombriser2017}. Why? Modern physics does not explain... 

In the end, it may turn out that these problems are not the problems of physics or cosmological constant, but the problem of understanding of our place in the world. In this context, the answer to these questions was proposed by S. Weinberg in 1987 on the basis of anthropic principle\footnote{It was first stated by B. Carter in 1973 at Krakow Symposium on occasion of the 400th anniversary of N. Copernicus birth \cite{Carter1974}.}, which can be formulated  briefly as follows: ''The Universe can not be another than it is, since we live in it``. In fact, if the cosmological constant had become only an order of magnitude larger than its observed value, then the Universe would have expanded so rapidly that the stars would not have been able to be formed, which means that the life would have not arisen.

Physicists, however, do not accept such answer to these questions unanimously, therefore, the searches are continuing. 
 
\section{Triumph of $\Lambda$}

The transition of observational astrophysics to digital light receivers, the launch of Hubble Space Telescope and the commissioning of ground-based telescopes with the mirror diameter 8 or more meters in the late 1980s created opportunities for implementation of the cosmological test ''apparent magnitude-redshift`` based on observations of the standardised light sources. Supernovae type Ia -- explosions of the white dwarfs in close binary systems -- appeared to be them. However, due to the rarity of such events -- several outbreaks in a thousand years in a galaxy -- and their short duration -- the duration of maximum brightness $\sim$5-10 days and $\sim$30-40 days for the entire period of visibility -- using them to obtain the results of required accuracy and statistical significance seemed unrealistic. In case of the successful implementation of such test one could determine the deceleration parameter of the Universe which in the notation of equations (\ref{fre}) is defined as follows:  $q\equiv-R\cdot R^{\prime\prime}/R^{\prime 2}$. If we introduce the dimensionless parameters of densities of the matter (in units of critical density) $\Omega_m\equiv8\pi G\rho_m(t_0)/3H_0^2$, the energy density of radiation $\Omega_{\gamma}\equiv8\pi G\rho_{\gamma}(t_0)/3H_0^2$, the 3-space curvature $\Omega_k\equiv\mathcal{K}c^2/H_0^2$ and the cosmological constant $\Omega_{\Lambda}\equiv\Lambda c^2/3H_0^2$, where $H_0=70h $ km/s$\cdot$Mpc is Hubble constant, then 
\begin{equation}
q=\frac{\Omega_{\gamma}(R_0/R)^4+\frac{1}{2}\Omega_m(R_0/R)^3-\Omega_{\Lambda}}
{\Omega_{\gamma}(R_0/R)^4+\Omega_m(R_0/R)^3+\Omega_k(R_0/R)^2+\Omega_{\Lambda}} \label{q}
\end{equation}
In the current epoch ($R=R_0$) the denominator is equal to 1, which follows from the second Friedmann equation (\ref{fre}): $\Omega_{\gamma}+\Omega_m+\Omega_k+\Omega_{\Lambda}=1$. Since the density of radiation is well-known -- in the current epoch $\Omega_{\gamma} = 2.49\cdot10^{- 5}h^{-2}$, this addition can now be neglected. Also at that time it was assumed that $\Lambda=0$, so the deceleration parameter in current epoch is defined as follows:  $q_0\approx\frac{1}{2}\Omega_m=\frac{1}{2}(1-\Omega_k)$. It is always positive, and its determination would give the values of average density of the Universe and curvature of 3-space -- basic parameters of any cosmological model. That is why the implementation of this test has become crucial for cosmology.  

In 1989 the project Calan/Tololo Supernova Survey \cite{Hamuy1993} began to operate.  Its purpose was to construct the Hubble diagram ''apparent magnitude-distance`` for supernovae type Ia, which are at the redshifts up to 0.1, that is up to distances $\sim$400 Mpc/h. Until 1993 the researchers discovered 38 supernovae type Ia and showed that their brightness could be standardised if their brightness curves, color indexes and spectra were measured.  
This became the basis for moving to larger redshifts. In 1994 the active phase of search for supernovae type Ia in distant galaxies was started independently by two large intercontinental teams Supernova Cosmology Project (SCP) and High-z Supernova Search (HzSS) which involved the most powerful observational resources of the ground-based telescopes and Hubble Space Telescope. The first team consisted of 33 scientists from the USA, Australia, France, Portugal, Spain, Great Britain and Chile, and Sol Perlmutter from Lawrence National Laboratory in Berkeley, USA was its principal investigator. The second one consisted of 26 members from five astronomical observatories and seven universities in the world, its leader was Brian Schmidt from Australian National University. By the year 1998 within the SCP project 79 supernovae \cite{Perlmutter2011} were discovered, among which 42 were supernovae type Ia \cite{Perlmutter1999}. Within the HzSS project 16 supernovae type Ia were discovered and studied in detail. Each team developed its own method of standardising the brightness of supernovae type Ia taking into account half-width of the light curve, color indices, features in the spectrum, evolution and absorption of light by the interstellar medium of parent galaxy etc. The main results  \cite{Schmidt1998,Riess1998,Perlmutter1999} following from the obtained relations ''apparent magnitude - redshift`` for standardised supernovae Ia were published practically simultaneously: in September-November 1998 and in June 1999 respectively. Main conclusion -- the Universe is expanding with the acceleration, which is caused by the positive cosmological constant. Both teams have obtained practically the same constraints on the relation between density of matter and density that corresponds to the cosmological constant: $0.8\Omega_m-0.6\Omega_{\Lambda}\approx-0.2\pm0.1$. From this relation it follows that $q_0\approx-0.83\Omega_m-0.33\pm0.17<0$ for all $\Omega_m>0 $. That is, the Universe expands with positive acceleration as in the models with positive cosmological constant. Its value depends on the curvature of 3-space which is not determined by these measurements. If it is zero (flat or Euclidean 3-space) then $\Omega_{\Lambda}\approx0.71\pm0.07$. The probability of zero hypothesis -- ''the value of cosmological constant is zero`` -- was only a few hundredths of a percent. The results became scientific sensation -- after 80 years from introduction of the cosmological constant ''into life`` by Einstein it finally got the reliable observational basis. In subsequent years observable data that confirmed the results of these two teams were obtained. As recognition of the importance of obtained results Sol Perlmutter, Brian Schmidt and Adam Riess were awarded the Nobel Prize in Physics\footnote{Another key member of both teams and co-author of all three Nobel papers, Alex Filippenko, professor at University of California, Berkeley, USA was awarded prestigious awards: Gruber Cosmology Prize (2007) and Breakthrough Prize in Fundamental Physics (2015).} in 2011. In their Nobel lectures \cite{Perlmutter2011,Schmidt2011,Riess2011} all three laureates have stressed unexpectedness of the result that supports existence of the cosmological constant. In particular, Brian Schmidt wrote: ''To my surprise, the accelerating Universe was received with a warmer reception than what I was expecting``. He explains this by the fact that two competing teams obtained independently the same results, which raised the level of trust to them. But there is also another explanation -- the cosmological community was expecting such a result, since in the mid-90s there were already indications that observable data on the large-scale structure of the Universe prefer the models with cosmological constant. The fact is that in 1992 the results of space experiment COBE (COsmic Background Explorer) on measurements of the CMB temperature fluctuations were published\footnote{The principal investigator of this part of the experiment George Smoot was awarded the Nobel Prize 
in physics in 2006.}: $<\left(\Delta T/T\right)^2>^{1/2}=(1.1\pm0.1)\cdot10^{-5}$ at angular scales $\geq10^{\circ}$. Since in the probable scenario of formation of the large-scale structure of the Universe from the primordial adiabatic density perturbations generated in inflation epoch the perturbations of CMB temperature and matter density are unambiguously related, these results allowed to establish the amplitude of power spectrum of the matter density perturbations at the scale of particle horizon at the time of decoupling of radiation from plasma at the cosmological recombination. Most models of the inflation predict scale-invariant primordial power spectrum of the adiabatic mode of matter scalar perturbations $\mathcal{P}_{pr}(k)=Ak^n$ with $n=1$, or close to it. The results of COBE and subsequent experiments have confirmed this prediction. For given composition of matter in the Universe and for known $\mathcal{P}_{pr}(k)$ one can calculate the evolution of initial perturbations from inflation to the modern epoch, the characteristics of large-scale structure of the Universe such as the spatial two-point correlation functions of galaxies, rich clusters of galaxies, the mass function of clusters of galaxies, the peculiar velocities of galaxies etc. Comparison of them with the corresponding observational data led to conclusion that the standard model with cold dark matter (sCDM: $\Omega_{cdm}=0.95$, $\Omega_b=0.05$, $H_0$=50 km/s$\cdot$Mpc, $n=1$) and normalised to the results of COBE initial spectrum of perturbations is not the model of observed Universe. This problem was most completely described in 1993 in the paper by Andrew Liddle and David Lyth \cite{Liddle1993}. Among the possible alternatives to sCDM the authors also mentioned the model with cosmological constant. The search for model consistent with all data of the observational cosmology became a key problem of cosmology of the 1990s. Most researchers came to conclusion that the model with cosmological constant is best fitting. Ricardo Valdarnini from International School for Advanced Studies (SISSA) in Trieste, Italy, Tinatin Kahihanishvili from Abastumani Astrophysical Observatory in Tbilisi, Georgia and I joined the discussion attempting to prove that the observed data on large-scale structure of the Universe can be made consistent with the COBE data within the model with mixed cold and hot dark matter without cosmological constant. We assumed three sorts of massive neutrinos as hot dark matter. It seemed to us then that such model is more promising, since it has more free parameters -- the total density and the ratios of neutrino masses -- than the model with cold dark matter and cosmological constant. However, very soon we became convinced that without cosmological constant we would not obtain the desired agreement of the model predictions with the observational data: the square of deviation ($\chi^2$) of the theoretically predicted characteristics from the corresponding observational ones decreased radically in the models with cosmological constant. Comparing the calculated power spectra of spatial inhomogeneities of matter, the root mean square of peculiar velocities of galaxies averaged at the scales of $\sim10-60$ Mpc, the two-point spatial correlation functions of the rich clusters of galaxies, their functions of mass and X-ray temperature with available observable data we came to the conclusion that in case of the Hubble constant equal to 70 km/c$\cdot$Mpc the density of hot component does not exceed 20\% of the total matter density (cold dark matter + hot dark matter + baryons), which is within the limits $0.3\le\Omega_m\le0.5$, and the cosmological constant in the same units of dimensionless density is within $0.5\le\Omega_{\Lambda}\le0.7$. The probability of null hypothesis was several percent. These results were reported by co-authors at various conferences in 1996-1999 years and published in their proceedings. The main paper was prepared for Astron. Astrophys., sent to the journal on December 22 1997 and published in the August issue 
\cite{Valdarnini1998}. In it we also mentioned the papers by many other authors who in those years discussed the models with nonzero $\Lambda$. Among them there is important paper of 1984 by Philip J. E. Pebbles \cite{Peebles1984}, which is still actively cited. Taking into account the data on peculiar velocities of galaxies caused by the local matter inhomogeneities, from which the mean value of matter density within $0.1\le\Omega_m\le0.3$ followed, age of the oldest stars and prediction of the flat 3-space by inflation models, he came to conclusion that the $\Lambda$CDM model is likely from the point of view of interpretation of the observed data. He also pointed out the problems of fine tuning and coincidence mentioned above.   

\section{Dark energy or cosmological constant?}

Long unsolvability of the problem of physical interpretation of the cosmological constant has led to the search for its alternatives which are actively developing both in terms of theoretical models and their testing by observational data. In 1998 Michael Turner called them the dark energy \cite{Turner1999a,Turner1999b}. This term combines various physical essences which in their manifestations are close to the cosmological constant, explain the accelerated expansion of the Universe, the observed large-scale structure of the Universe, its composition and the CMB anisotropy. They can be divided into two broad classes: the fields which homogeneously fill the Universe and the modifications of gravity. Each one has too many models to describe them here. It is important that a large number of ''viable`` models are indistinguishable at the current level of accuracy of cosmological observations: they are equally consistent with all available observable data and nobody sees the possibility of their identification with key experimental tests. At the same time, the future destinies of the Universe predicted by them are different: eternal exponential expansion, as in de Sitter model, eternal power-law expansion or turn around and collapse to the singularity Big Crunch in the quintessential dark energy model, superfast expansion to the singularity Big Rip in the phantom model of dark energy. In different models with the same age of the Universe today -- 13.7$\pm$0.2 billion years -- the future duration of observed evolution is different. That is, if in the 1980s we could not predict future of the Universe because of the insufficient number and accuracy of the observational data, now we are not able to do this because of the excessive number of theoretical models which cannot be distinguished by observations. And it is unknown whether such an opportunity will be in principle. It may turn out that the dark sector has as many ''things`` as has the ''light`` one which is described by the standard model of particle physics. In 1998 Michael Turner named two of his papers optimistically ''Cosmology Solved? Maybe'' and ''Cosmology Solved? Quite Possibly! `` \cite{Turner1999a,Turner1999b}. However, after eight years of active studies of the dark energy models another well-known astrophysicist Thanu Padmanabhan answered him by the less optimistic title: ''Dark Energy: Mystery of the Millennium`` \cite{Padmanabhan2006}. In my opinion, the present state of understanding of the nature of dark energy is reflected by the title of article by Vladimir Burdyuzha ''The Dark Components of the Universe Are Slowly Clarified`` \cite{Burdyuzha2017}. 
 
\section{Acknowledgements} 
This work was supported by the project of Ministry of Education and Science of Ukraine (state registration number 0116U001544). Author is thankful to academician Valerij Shulga and professor Han Wey for invitation to International Center of Future Science of Jilin University (P.R.China) where this paper has been finished.

\end{document}